\documentclass[a4paper]{spie}  
\usepackage[]{graphicx}
\usepackage{amssymb,amsmath,psfrag,url}

\title{Finite-Element Method Simulations of High-Q Nanocavities with 1D Photonic Bandgap}

\author{
Sven Burger,\supit{\,ab}
Jan Pomplun\supit{\,b}
Frank Schmidt\supit{\,ab}
Lin Zschiedrich,\supit{\,b}
\skiplinehalf
\supit{a}
Zuse Institute Berlin (ZIB),
Takustra{\ss}e 7,
D\,--\,14\,195 Berlin,
Germany
\smallskip\\
\supit{b}
JCMwave GmbH,
Bolivarallee 22, 
D\,--\,14\,050 Berlin,
Germany
}
\authorinfo{
Corresponding author: S. Burger\\
URL: http://www.zib.de/en/numerik/computational-nano-optics.html\\
URL: http://www.jcmwave.com\\
Email: burger@zib.de
}

\begin{document}
\maketitle
\noindent
This paper will be published in Proc.~SPIE Vol. {\bf 7933}
(2011) 79330T 
({\it Physics and Simulation of Optoelectronic Devices XIX, Bernd Witzigmann; Fritz Henneberger; Yasuhiko Arakawa; Alexandre Freundlich, Editors}, 
DOI: 10.1117/12.875044),
and is made available 
as an electronic preprint with permission of SPIE. 
One print or electronic copy may be made for personal use only. 
Systematic or multiple reproduction, distribution to multiple 
locations via electronic or other means, duplication of any 
material in this paper for a fee or for commercial purposes, 
or modification of the content of the paper are prohibited.

\begin{abstract}
High-Q optical resonances in photonic microcavities 
are investigated numerically 
using a time-harmonic finite-element method.  
\end{abstract}


\section{Introduction}

Optical microcavities allow to confine light to small volumes. 
High resonance Q-factors can be attained using the high reflectivity of multi-layer 
Fabry-Perot resonators, of total internal reflection and/or of photonic 
bandgap materials~\cite{Vahala2003a,Notomi2010rep}.
In this contribution we revisit a microcavity design proposed originally by Notomi, Kuramochi and Taniyama~\cite{Notomi2008oe}.
Notomi {\it et al} have shown that very high Q-factors can be reached with a size-modulated 
1D stack design. 
In this case confinement to the cavity is obtained by total internal reflection in two dimensions, and by a photonic 
bandgap in the third dimension. 
Advantages of this design are its compactness and simplicity. 

For an efficient design of microcavities and other integrated photonic components 
3D simulations of Maxwell's equations are 
needed. We have developed finite-element method (FEM) based solvers for the 
Maxwell eigenvalue and for the Maxwell scattering problems. 
The method is based on higher order vectorial elements, 
adaptive unstructured grids, and on a rigorous treatment of 
transparent boundaries.
We perform a numerical analysis of the microcavity setup. Results on resonance wavelengths and Q-factors are generated using a 
resonance solver. The obtained values are confirmed by simulations of  transmission spectra of light incident to the 
microcavities. 
Dependence of Q-factors and resonance wavelength on structural parameters are investigated. 
A focus of this contribution is on the numerical convergence of the obtained results. 

\begin{figure}[b]
\begin{center}
\psfrag{Ni}{\sffamily $N_i$}
\psfrag{No}{\sffamily $N_o$}
\psfrag{outgoing waveguide}{\sffamily outgoing waveguide}
\psfrag{incoming waveguide}{\sffamily incoming waveguide}
\psfrag{central block}{\sffamily central block}
\psfrag{Wz0}{\sffamily $W_{z,0}$}
\psfrag{Wzi}{\sffamily $W_{z,n}$}
\psfrag{WzN}{\sffamily $W_{z,N_i}$}
\psfrag{i=1}{\sffamily \tiny $i=1$}
\psfrag{i=2}{\sffamily \tiny $i=2$}
\psfrag{i=3}{\sffamily \tiny $i=3$}
\psfrag{x}{\sffamily x}
\psfrag{y}{\sffamily y}
\psfrag{z}{\sffamily z}

  \includegraphics[width=.98\textwidth]{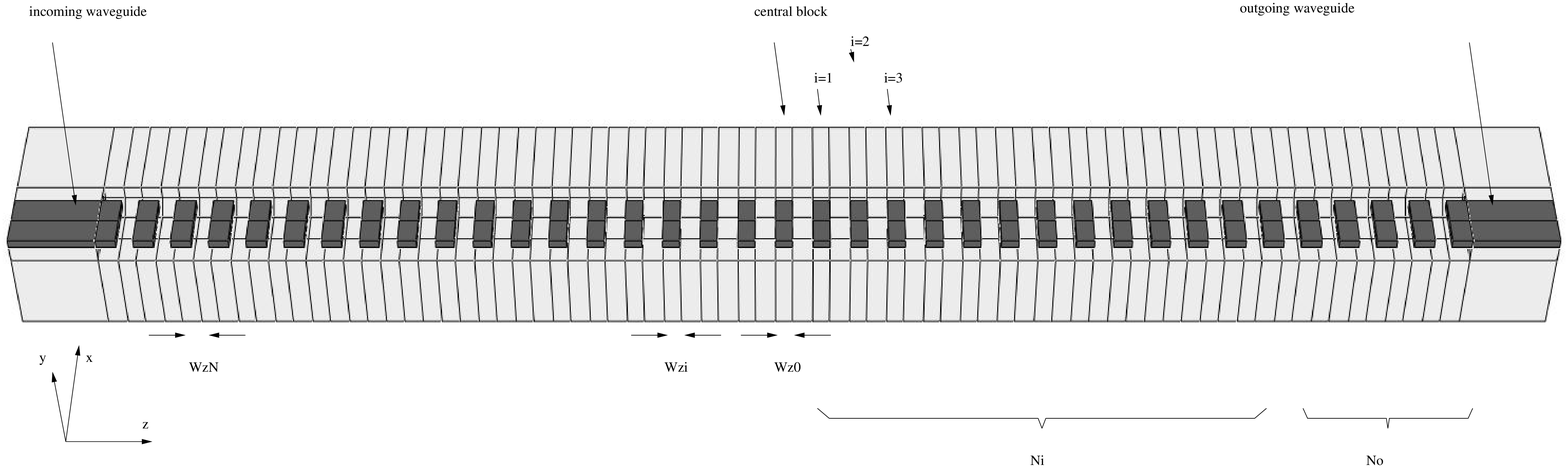}
  \includegraphics[width=.98\textwidth]{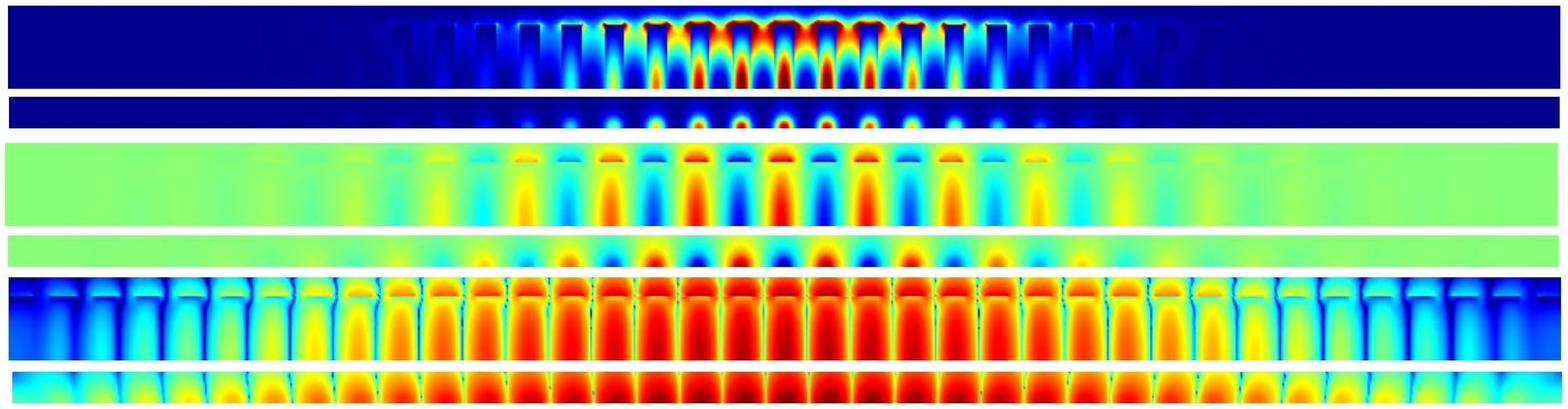}
  \caption{
Top:
Geometry of a size-modulated 1D stack microcavity
with a central block of width $W_{z,0}$, $2\times N_i=2\times 13$ size-modulated blocks of 
widths  $W_{z,n}$ and $2\times N_o=2\times 5$ blocks of constant width $W_{z,N_i}$. 
Blocks and incoming/outgoing waveguides consist of dielectric material, the 
structure is surrounded by air, waveguides extend to infinity. 
The structure is mirror-symmetric at the planes $x=0$, $y=0$, therefore the computational domain 
can be reduced to one quarter of the whole structure. 
Bottom: Field distribution on resonance (pseudo-color representations), from top to bottom:
intensity in a $x$-$z$-cross-section, intensity in a $y$-$z$-cross-section, 
$\Re(E_x)$ in a $x$-$z$-cross-section, $\Re(E_x)$ in a $y$-$z$-cross-section, 
$\log(E_x^2)$ in a $x$-$z$-cross-section, $\log(E_x^2)$ in a $y$-$z$-cross-section, 
}
\label{fig_grid}
\end{center}
\end{figure}

\section{Investigated Setup}
\label{section_setup}
The investigated setup consists of a finite array of dielectric blocks, as depicted in 
Figure~\ref{fig_grid}. The first and the last 
block of the array are merged to waveguides with same square shape as the blocks, but infinite 
in the third dimension ($z$-direction). 
The centers of the blocks are arranged periodically with a period of $a = 430\,$nm.  
Width and height of the waveguide and blocks (in $x$- and $y$-direction) are $W_x=3a$ and $W_y=0.5a$.
The length of $2 N_i +1$ inner blocks is modulated symmetrically around a central block, 
$W_{z,n} = W_{z,0}+(|n|/N_i)^2(W_{z,N_i}-W_{z,0})$, with $W_{z,0}=0.45a$,  $W_{z,N_i}=0.5345\,a$. 
The length of $2N_o$ outer blocks is constant $W_{z,n}=W_{z,N_i}$ ($n>N_i$), 
where the distance from the center block is denoted with $n$ (in number of blocks). 
The setup of Notomi {\it et al} corresponds to $N_i=13$.
The blocks consist of dielectric with refractive index $n=3.46$, the surrounding is filled with 
vacuum ($n=1.0$). 
Conceptually similar devices have also been investigated experimentally~\cite{Kuramochi2010oe}.

\section{Numerical Results}
\subsection{Band diagram of the 1D periodic array}

\begin{figure}[t]
\begin{center}
\psfrag{$\omega$}{\sffamily $\tilde{\omega}$}

  \includegraphics[width=.48\textwidth]{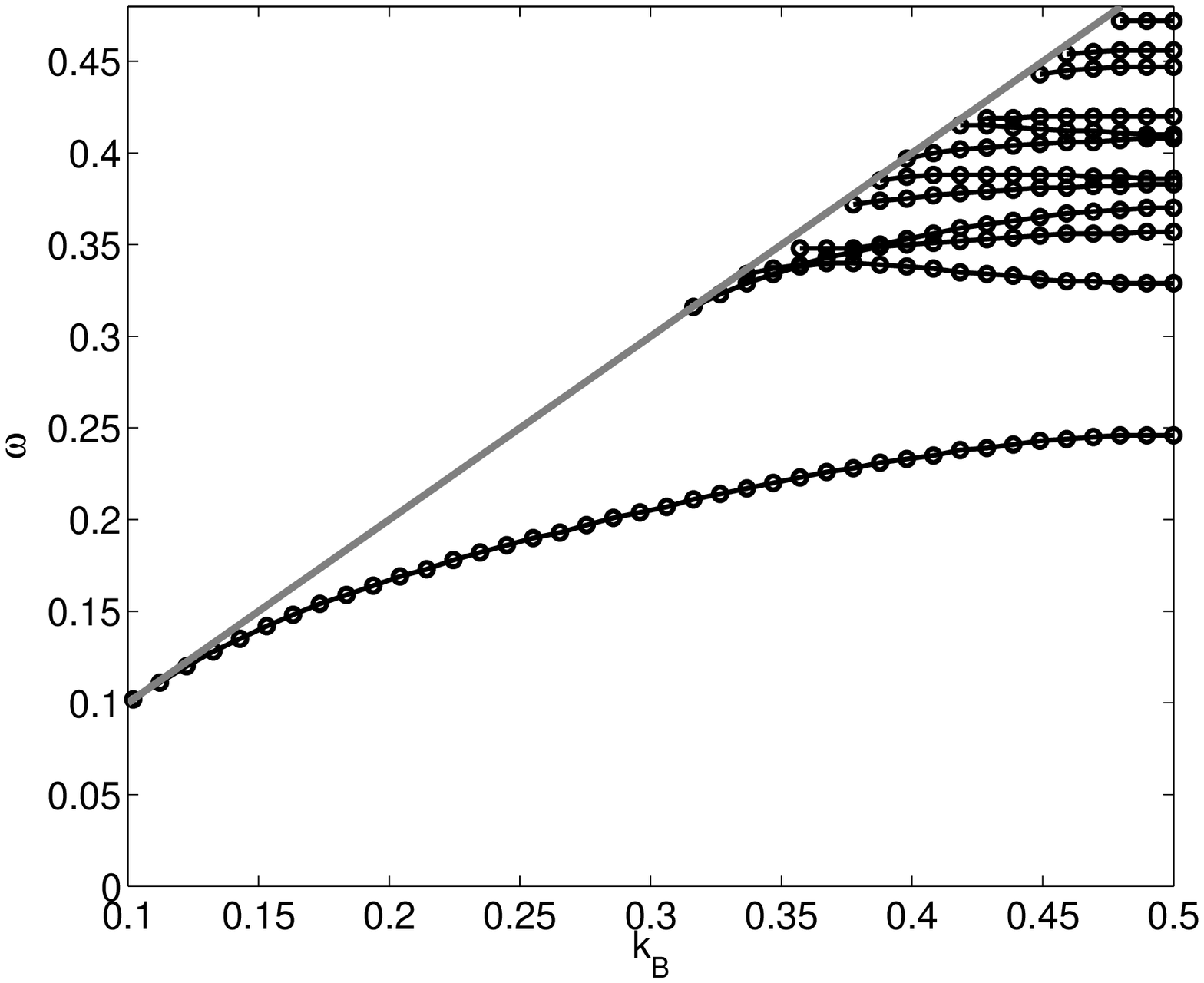}
  \includegraphics[width=.48\textwidth]{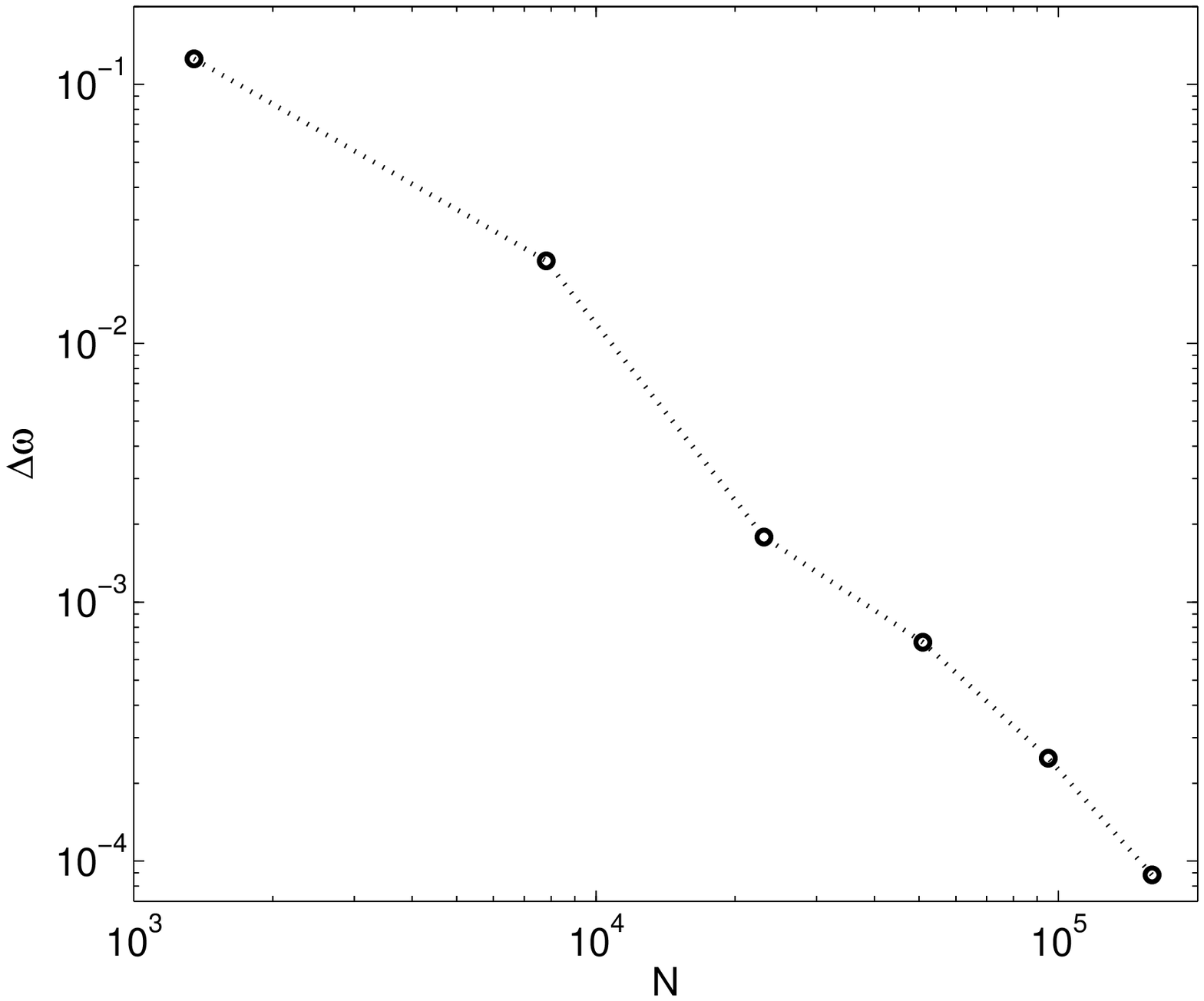}
  \caption{
Left: Part of the band structure of the 1D photonic crystal. 
The light cone  is indicated by a grey line. 
Right: Convergence of the relative error of the computed eigenfrequency of the lowest 
Bloch mode at $\tilde{k} = 0.5$.
}
\label{fig_band_diagram_1d}
\end{center}
\end{figure}

We start the numerical investigation by computing the band diagram of the 1D periodic array of dielectric 
blocks in vacuum. The unit cell of for the computation is periodic in $z$-direction with a period $a$, 
transparent boundary conditions (perfectly matched layers, PML) apply in $x$- and $y$-directions,
the width of the blocks in z-direction is $W_{z}=0.5345\,a$. 
Computation of modes with even symmetry is selected by reducing the symmetric computational domain to the half space $x>0$ 
and applying corresponding boundary conditions at the symmetry plane. 
Figure~\ref{fig_band_diagram_1d} (left) shows a part of the band structure of the 1D photonic 
crystal~\cite{Joannopoulos1995a}. 
A band gap appears for dimensionless frequencies between 
$\tilde{\omega}\approx 0.246$ and $\tilde{\omega}\approx 0.329$. 
Dimensionless frequency $\tilde{\omega}$ is defined as $\tilde{\omega} = (\omega a)/(2\pi c_0)$, with 
time-harmonic frequency $\omega$ and vacuum speed of light $c_0$. 
Dimensionless magnitude of Bloch vector $\tilde{k}$ is defined as $\tilde{k} = ka/(2\pi)$, with Bloch vector $k$.
For $a = 430\,$nm the bandgap corresponds to a wavelength range between approximately 1307\,nm and 1748\,nm.
In Figure~\ref{fig_band_diagram_1d} (left) only modes below the light cone are displayed. Modes above the light 
cone have complex eigenfrequencies, which corresponds to leakage to the transversal directions~\cite{Joannopoulos1995a}. 

Figure~\ref{fig_band_diagram_1d} (right) shows numerical convergence of the simulated lowest Bloch-mode at 
$\tilde{k} = 0.5$: 
Finite-element simulations have been performed for the same physical parameters and for different numerical parameters with 
increasing numerical accuracy. In this case the polynomial degree of the finite-element ansatzfunctions~\cite{Pomplun2007pssb}, 
$p$, has been varied between $p=1$ and $p=7$ for a fixed spatial mesh discretizing the geometry of the unit cell. 
The computed value of  $\tilde{\omega}$ for $p=7$,  $\tilde{\omega}_{p=7}\approx 0.24644$, is taken as quasi-exact value, and the 
relative error of the computed values of $\tilde{\omega}$ for $p=1\dots 6$, 
$\Delta\tilde{\omega}_{p}=|\tilde{\omega}_{p}-\tilde{\omega}_{p=7}|/\tilde{\omega}_{p=7}$ is plotted as function of the number 
of unknowns of the discrete problem. 
Computation times on a standard workstation range between below a second and few minutes for the plotted values. 
High accuracy with relative errors better than 0.1\% is obtained for $p\ge 3$. 
The solvers used for the band diagram simulations and for the further simulations throughout this contribution 
are part of the FEM program package {\it JCMsuite}~\cite{Burger2008ipnra} which is developed 
by Zuse Institute Berlin and JCMwave. 

\subsection{Transmission of waveguide modes through a finite 1D periodic array}
\label{sec_phc_transmission}

\begin{figure}[b]
\begin{center}
  \includegraphics[width=.98\textwidth]{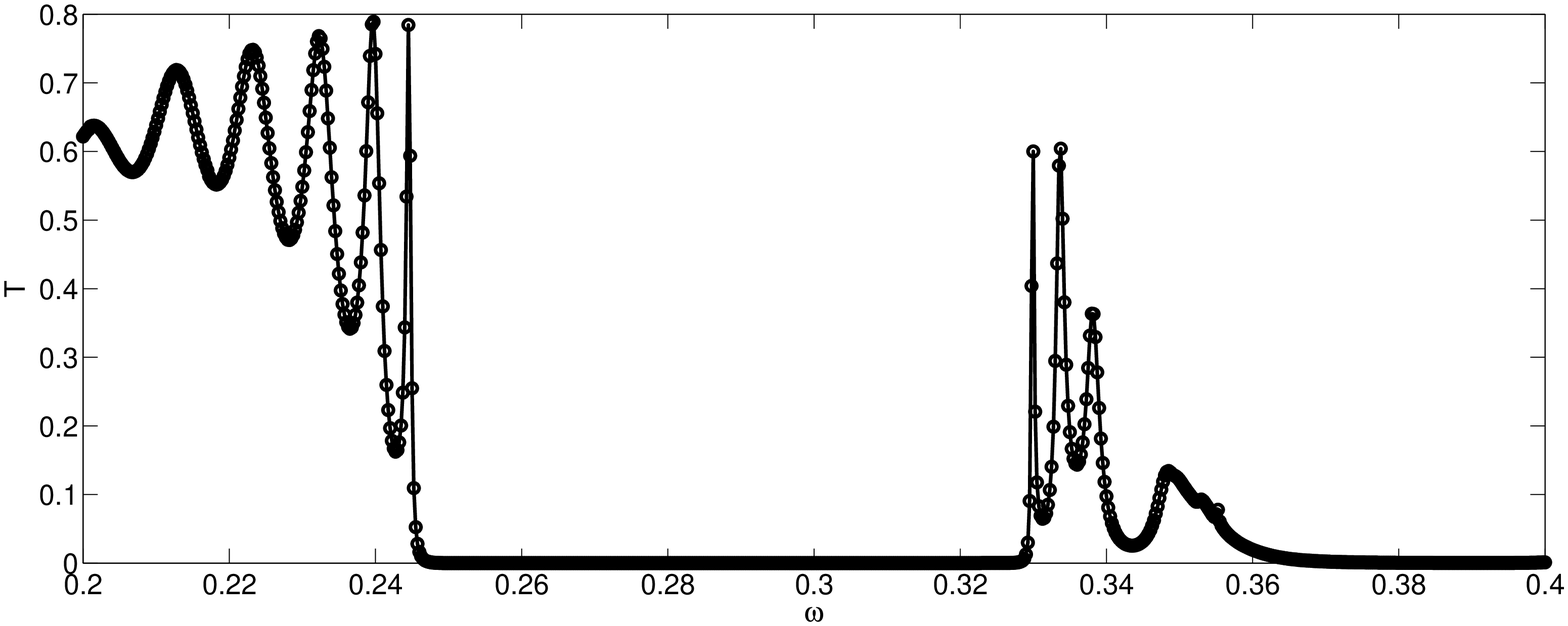}
  \caption{
Transmission spectrum of incident waveguide modes through a finite 1D periodic array of 15 blocks of length $W_{z}=0.5345\,a$.
Transmission is suppressed in the frequency range between $\tilde{\omega}\approx 0.246$ and $\tilde{\omega}\approx 0.329$.
}
\label{fig_transmission_1d}
\end{center}
\end{figure}

For simulating transmission through a finite array of blocks / a finite 1D photonic crystal 
at specific wavelength $\lambda_0$, 
we first compute the fundamental 
propagation mode of the waveguide at $\lambda_0$ using a FEM  
propagation mode-solver. 
The obtained mode field is applied as input data to one of the boundaries of the 3D 
computational domain (left boundary in Fig.~\ref{fig_grid}), such that the mode propagates 
in direction of the center of the waveguide. 
We then compute the scattered light field in the setup corresponding to this excitation 
using higher-order finite-elements. 
In post-processes we extract energy fluxes through interfaces and field distributions 
in several cross-sections from the 3D light field distribution.
Transmission $T$ is defined as ratio of the energy flux of the outgoing light field at the back side of the array
to the energy flux of the incoming waveguide mode through the 
waveguide cross section. 

Figure~\ref{fig_transmission_1d} shows a transmission spectrum for 800 incident waveguide modes with vacuum 
wavelengths between $\lambda_0=1075\,$nm ($\tilde{\omega}=0.4$) and  $\lambda_0=2150\,$nm ($\tilde{\omega}=0.2$).
In perfect agreement with the band structure simulations transmission is greatly suppressed 
in the frequency range between $\tilde{\omega}\approx 0.246$ and $\tilde{\omega}\approx 0.329$
which corresponds to a bandgap of the 1D periodic structure ({\it cf}~Figure~\ref{fig_band_diagram_1d}).

\begin{figure}[b]
\begin{center}
\psfrag{Ni}{\sffamily $N_i$}
\psfrag{N i}{\sffamily $N_i$}
\psfrag{L [nm]}{\sffamily $\lambda_0$ [nm]}
\psfrag{Q}{\sffamily Q-factor}
  \includegraphics[width=.48\textwidth]{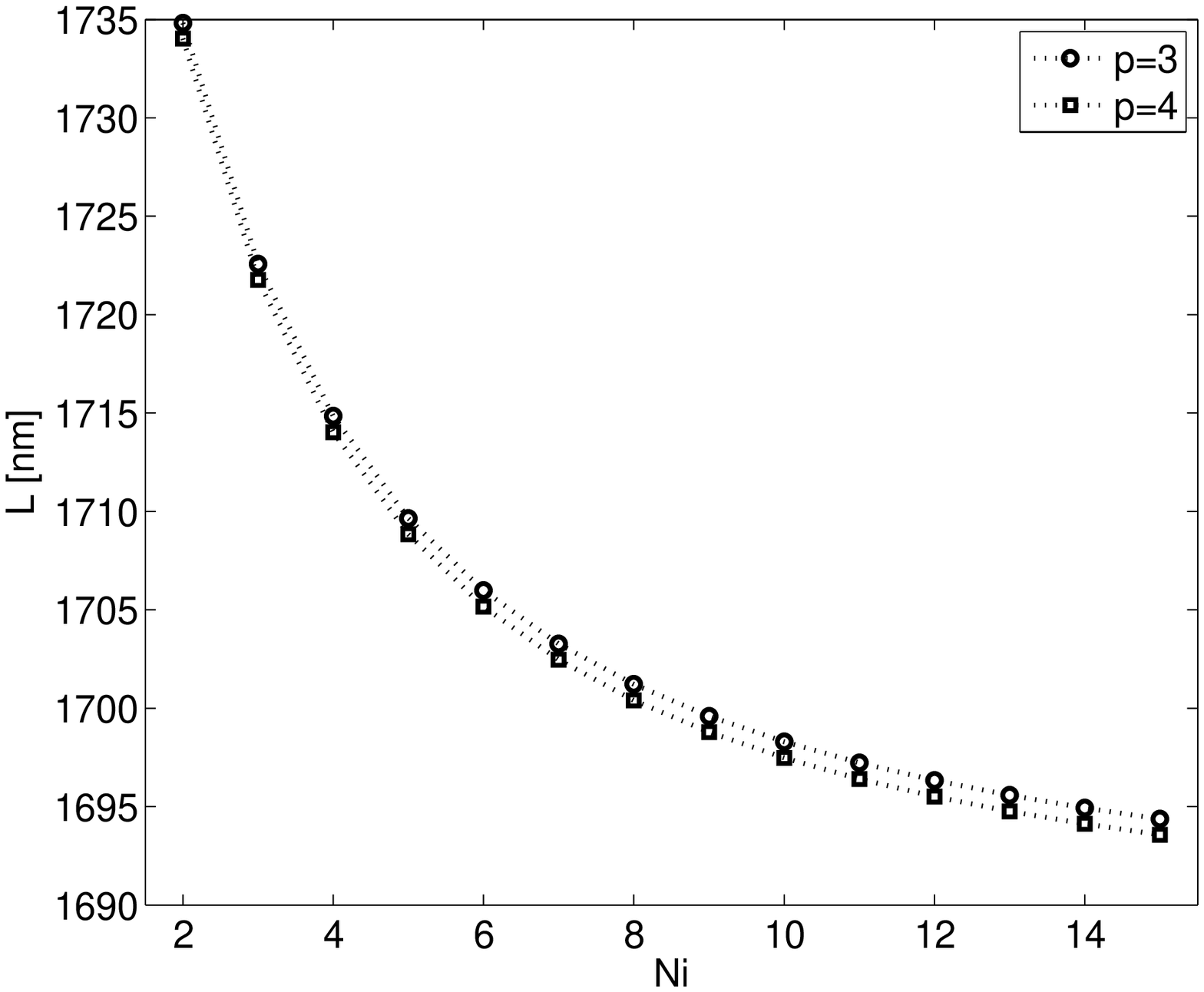}
  \includegraphics[width=.48\textwidth]{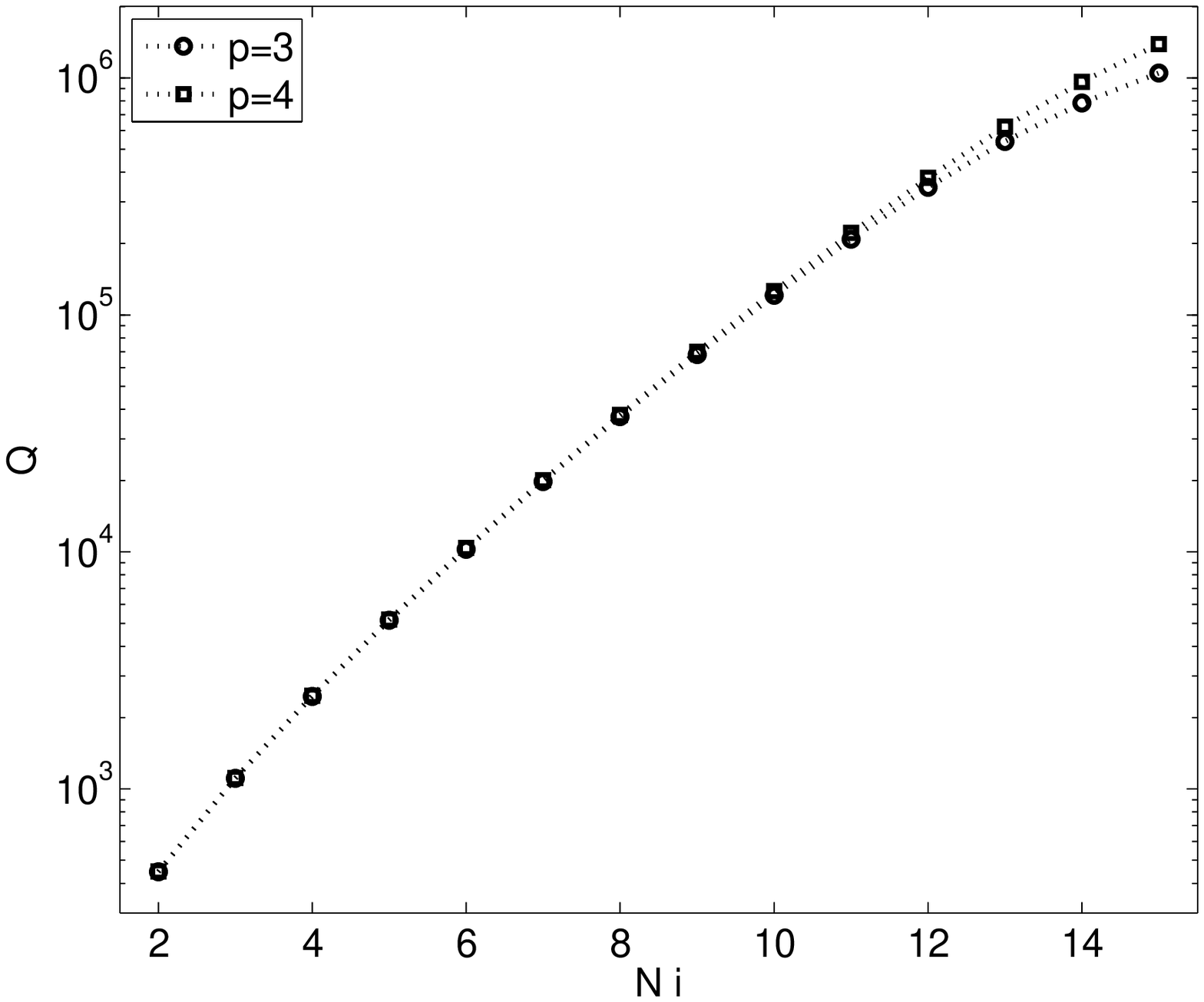}
  \caption{
Dependence of resonance wavelength $\lambda_0$ (left) and Q-factor (right) on 
number of modulated blocks, $N_i$. For all simulations, $N_o = 4$.
Results are displayed for finite-element degree $p=3$ and  $p=4$. 
}
\label{fig_resonance_ni}
\end{center}
\end{figure}

\subsection{Direct simulation of high-Q cavity resonances}
\label{sec_resonances}
 
We use an eigenmode solver for directly simulating the resonance properties (resonance wavelength 
and Q-factor) of size-modulated 1D stack cavities: 
Given the geometrical setup as described in Section~\ref{section_setup}, 
one computes an electric field distribution $E$
and a complex eigenfrequency $\omega$ which satisfy Maxwell's time-harmonic wave equation without sources,
$$\nabla \times \mu^{-1} \nabla \times E = \omega^2 \varepsilon E\quad ,$$
electric permittivity and magnetic permeability are denoted by $\varepsilon$ and $\mu$, respectively.
Transparent boundary conditions (PML) take into account the specific geometry 
of the problem where waveguides are modelled to  
extend to infinity in the exterior domain. 
When the eigenmode ($E, \omega$) is computed, the respective $Q$-factor is deduced from the real and imaginary parts 
of the complex eigenfrequency,
$Q=\Re(\omega)/(-2\Im(\omega)),$
the resonance wavelength $\lambda_{0}$ is given by $\lambda_{0}=2\pi c_0/\Re(\omega)$.
Figure~\ref{fig_grid} shows visualizations of a typical field distribtion obtained with the resonance solver. 

\begin{figure}[t]
\begin{center}
\psfrag{No}{\sffamily $N_o$}
\psfrag{N o}{\sffamily $N_o$}
\psfrag{Delta L [nm]}{\sffamily $\Delta\,\lambda_0$ [nm]}
\psfrag{Q}{\sffamily Q-factor}
  \includegraphics[width=.48\textwidth]{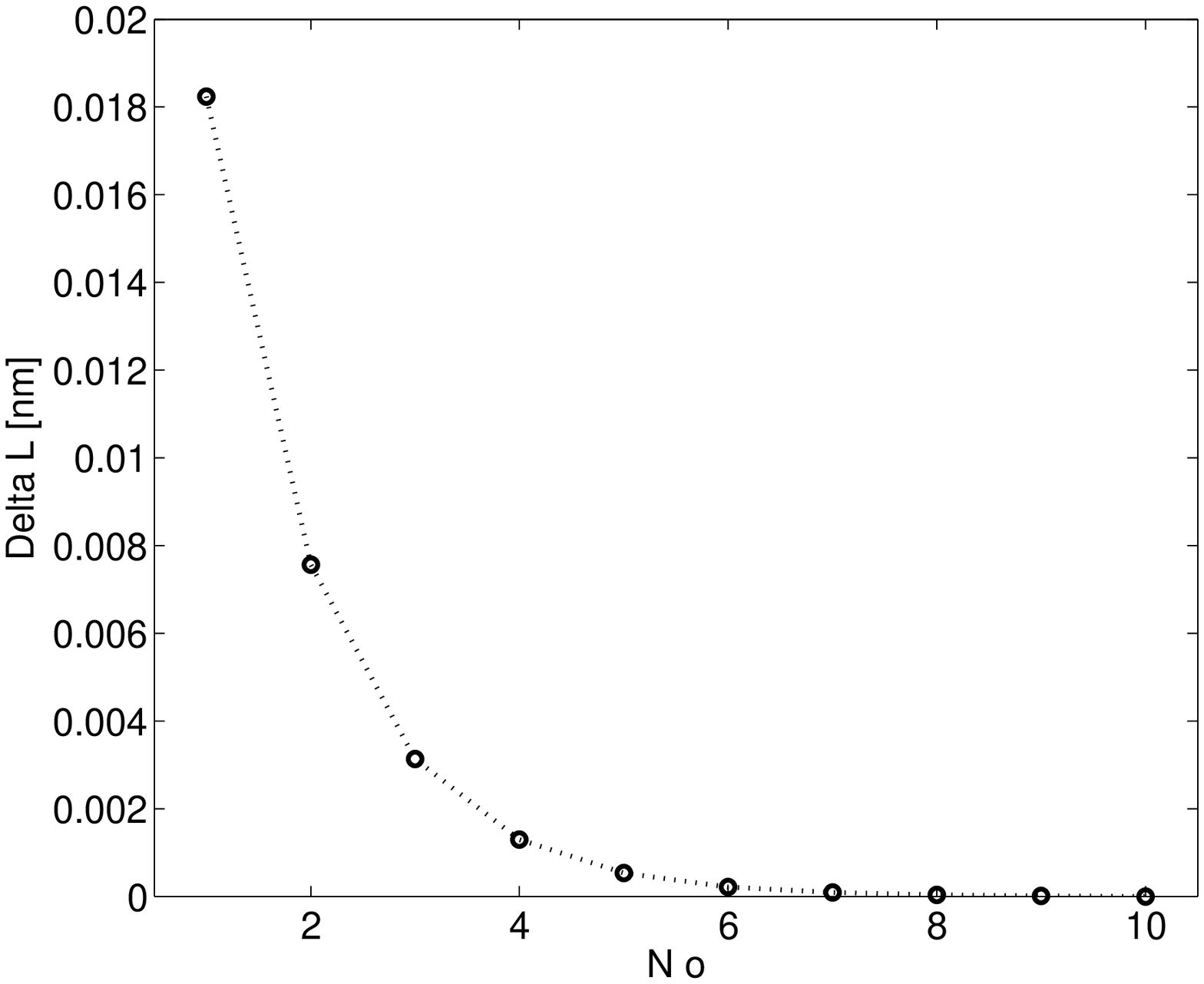}
  \includegraphics[width=.48\textwidth]{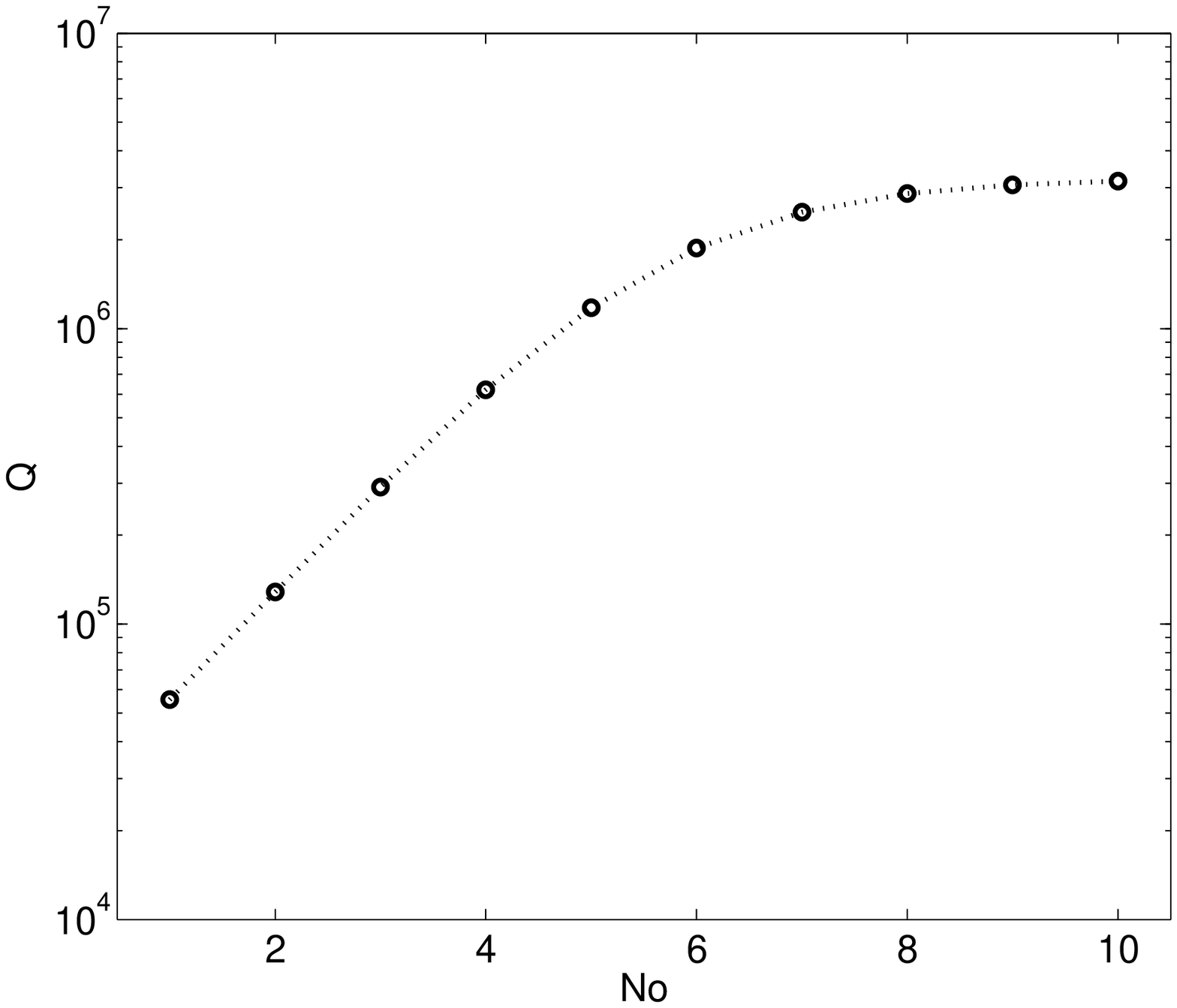}
  \caption{
Dependence of resonance wavelength $\lambda_0$ (left) and Q-factor (right) on 
number of outer, unmodulated blocks, $N_o$.
}
\label{fig_resonance_no}
\end{center}
\end{figure}

Direct simulation of a resonance requires a single computation only. 
In contrast, deducing the resonance properties from a transmission or 
reflection spectrum generated with a time-harmonic solver 
requires several computations at various wavelengths. 
Especially for high-Q resonances, where the choice of wavelengths for a transmission scan is not 
clear a-priori, direct computation of resonances simplifies the simulation task and reduces 
computational effort. 
Time-domain solvers for simulating 3D high-Q resonances typically suffer from very long computation times and
slow convergence rates. 

We have performed simulations for different cavity setups. 
Figure~\ref{fig_resonance_ni} shows the dependence of Q-factor and resonance wavelength on the number of 
modulated blocks, $N_i$, of the 1D stack cavity setup (see Section~\ref{section_setup}). 
For these simulations, the number of outer blocks is fixed to $N_o = 4$.
With increasing number of modulated blocks, the resonance wavelength is pulled from the band-edge (at 
a wavelength of 1748\,nm) more and more inside the band-gap. 
At the same time, the Q-factor increases nearly exponentially for the investigated range of $N_i=1\dots 15$.
Figure~\ref{fig_resonance_ni} shows results for different numerical resolutions, i.e., for finite-element degree 
$p=3$ and $p=4$. 
As can be seen from these results, 
for higher numbers of modulated blocks, i.e., for higher Q-factors, influence of numerical resolution on the results gets more significant,
up to a level of relative deviations of the order of ten percent. 

Figure~\ref{fig_resonance_no} shows the dependence of Q-factor and resonance wavelength on the number of 
outer, unmodulated blocks, $N_o$, of the 1D stack cavity setup (see Section~\ref{section_setup}). 
For these simulations, the number of inner blocks is fixed to $N_i = 13$ which corresponds to the setting of 
Notomi {\it et al}~\cite{Notomi2008oe}. 
For these simulations, fourth-order finite-elements ($p=4$) have been used. 
With increasing number of unmodulated blocks, the resonance wavelength is changed only on a sub-nanometer scale. 
The Q-factor increases exponentially for $N_o<6$ and reaches a plateau of $Q\approx 3\times 10^6$  for $N_o>7$. 
We expect that radiation losses to the sourrounding air limit high-Q performance in this case~\cite{Burger2010pw3}.

\begin{figure}[t]
\begin{center}
\psfrag{No}{\sffamily $N_o$}
\psfrag{N o}{\sffamily $N_o$}
\psfrag{Delta L [nm]}{\sffamily $\Delta\,\lambda_0$ [nm]}
\psfrag{Q}{\sffamily Q-factor}
  \includegraphics[width=.48\textwidth]{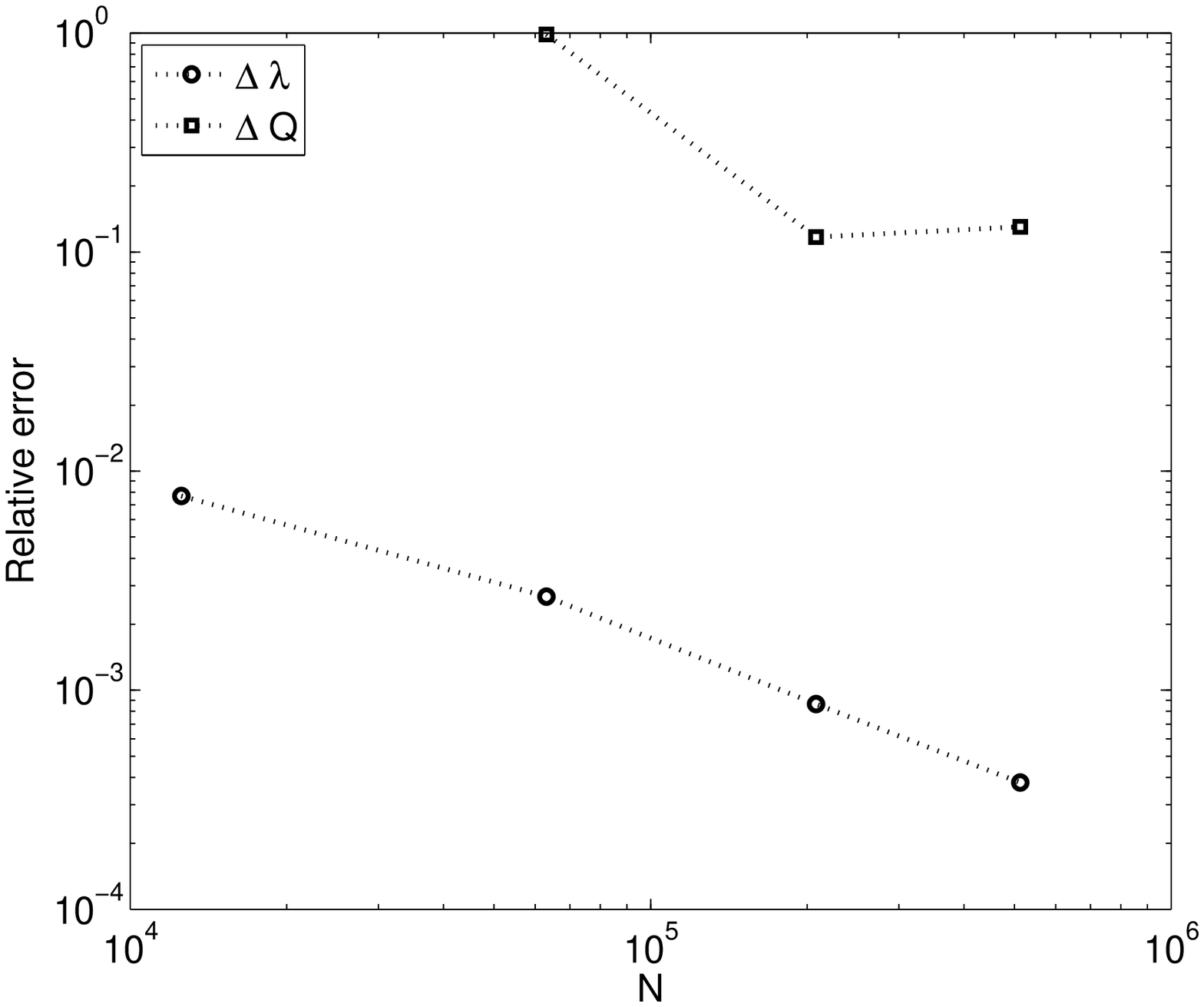}
  \includegraphics[width=.48\textwidth]{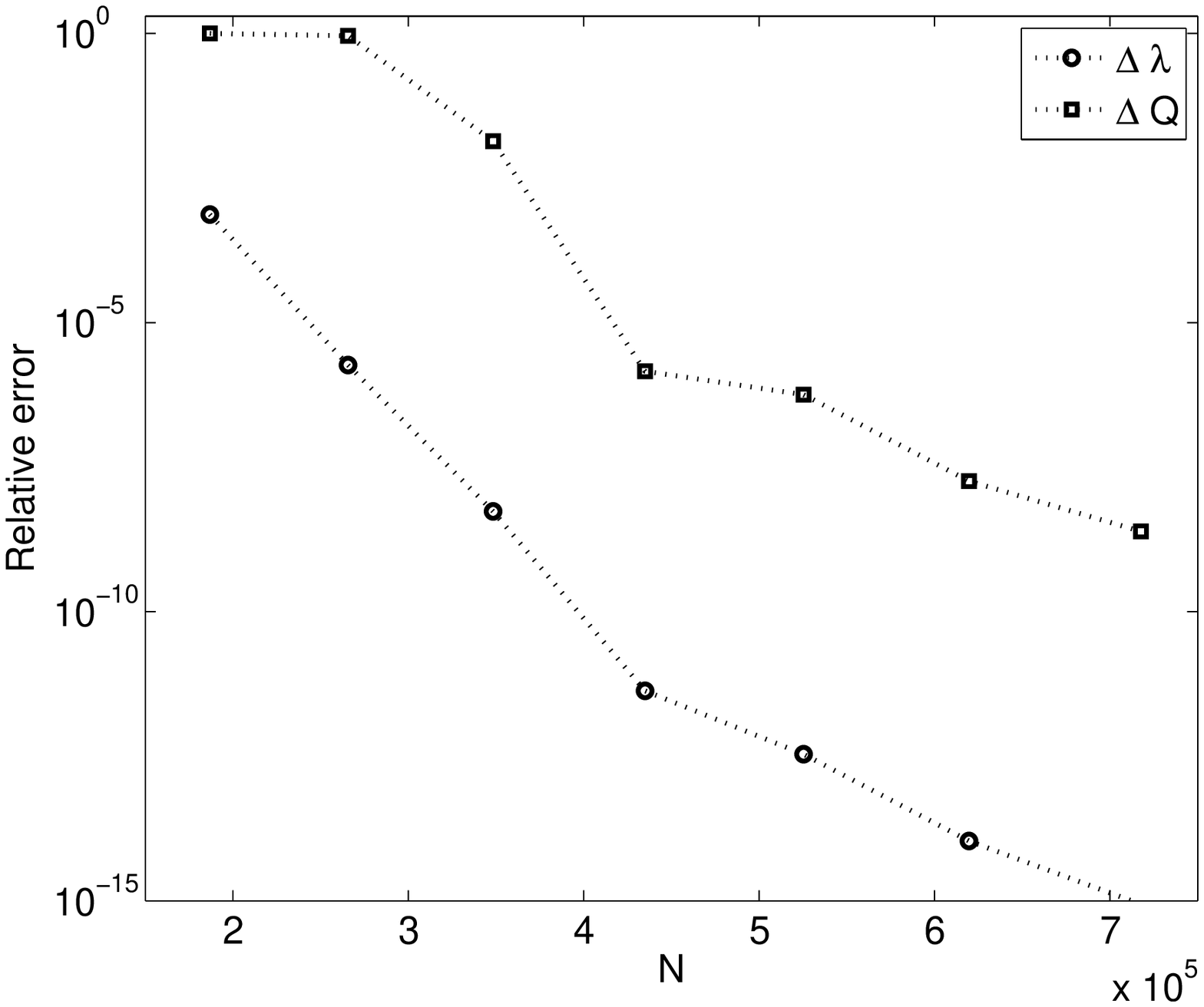}
  \caption{
Convergence of numerical results for high-Q ($10^6$) resonance simulation using a resonance solver:
Dependence of relative errors of Q-factor and resonance wavelength, $\Delta Q = |Q_N-Q_{qe}|/Q_{qe}$ and 
$\Delta \lambda = |\lambda_N-\lambda_{qe}|/\lambda_{qe}$, on number of unknowns of the FEM 
problem, $N$.
Geometry setting: $N_i = 13$, $N_o = 5$.  
Left: Finite-element degree $p$ varied between $p=1$ and $p=4$. 
Right: Number of segments discretizing the solution in the PML region varied between 1~and~7. 
}
\label{fig_resonance_conv}
\end{center}
\end{figure}

For evaluating numerical errors of high-Q resonance simulations using the resonance mode solver we again show
a numerical convergence study: 
The the geometry setting is fixed to to $N_i = 13$, $N_o = 5$.  
Finite-element polynomial degree $p$ is varied between $p=1$ and $p=5$, 
and  Q-factor,  resonance wavelength and numerical effort are recorded. 
Figure~\ref{fig_resonance_conv} (left) shows how the relative error of Q-factor and resonance wavelength converge with 
number of unknowns $N$, where the solution at highest resolution ($p=5$) has been taken as reference 
result ($\lambda_{qe}=1694.12\,$nm, $Q_{qe}=1.04\times10^6$, $N_{qe}=978,570$). 
Computation times for the displayed results are 1\,sec, 12\,sec, 2\,min, 9\,min, respectively (on a workstation 
with 8 CPU cores). 
This demonstrates that resonances with Q-factors of about 1\,million can be computed at accuracies of 
about 0.1\,\% (resonance wavelength) and 10\,\% (Q) within few minutes. 

In principle, a wrong numerical realization of transparent boundary conditions can introduce errors which will 
{\it not} show up in a convergence study as shown in Figure~\ref{fig_resonance_conv} (left).
We realize transparent boundary conditions with the PML method, 
i.e., by a coordinate transform to complex space, by FEM discretization with higher-order elements
(same order $p$ as in the interior domain), and by truncation at a variable distance from the boundary to the 
interior computational domain. 
For a fixed setting of $p=4$ we have further investigated dependence of the results on varied number of 
PML segments. 
Figure~\ref{fig_resonance_conv} (right) shows that convergence of the results with PML discretization parameter is 
very fast. This indicates that the main contribution to the numerical errors as shown in Figure~\ref{fig_resonance_conv} (left)
are caused by discretization of the electromagnetic field in the interior domain. 
Typical settings for the results displayed in Figure~\ref{fig_resonance_conv} (left) and 
Figure~\ref{fig_resonance_ni},~\ref{fig_resonance_no}
 are about seven PML segments. 
The setting of further PML parameters is done adaptively~\cite{Zschiedrich2006a}. 
More detailed studies should also consider influence of these parameters on numerical accuracy. 

We note that an evaluation of the significant differences between the numerical 
results of Notomi {\it et al}~\cite{Notomi2008oe} and our numerical results requires further investigations. 
Notomi~{\it et al} have reported results from a finite-difference time-domain method 
yielding significantly higher Q-factors (by about two orders of magnitude) for the same physical setting.

\subsection{Light scattering response of a high-Q cavity}
\label{sec_cavity_transmission}

An alternative possibility of computing the resonance properties of a cavity is to investigate its scattering response 
to incident light fields. 
Using the same approach as in Section~\ref{sec_phc_transmission} we perform FEM light scattering simulations where 
the incident light fields are waveguide modes at fixed frequencies/wavelengths. 
From the near field solutions we deduce the energy stored in the central part of the cavity, and the energy flux 
transmitted through the device. 

\begin{figure}[t]
\begin{center}
\psfrag{Wavelength [nm]}{\sffamily $\lambda_{in}$\,[nm]}
\psfrag{Stored Energy (normalized)}{\sffamily \hspace{2cm} $E$}
\psfrag{Transmission}{\sffamily \hspace{0.5cm} $T$}
\psfrag{Q}{\sffamily Q-factor}
  \includegraphics[width=.48\textwidth]{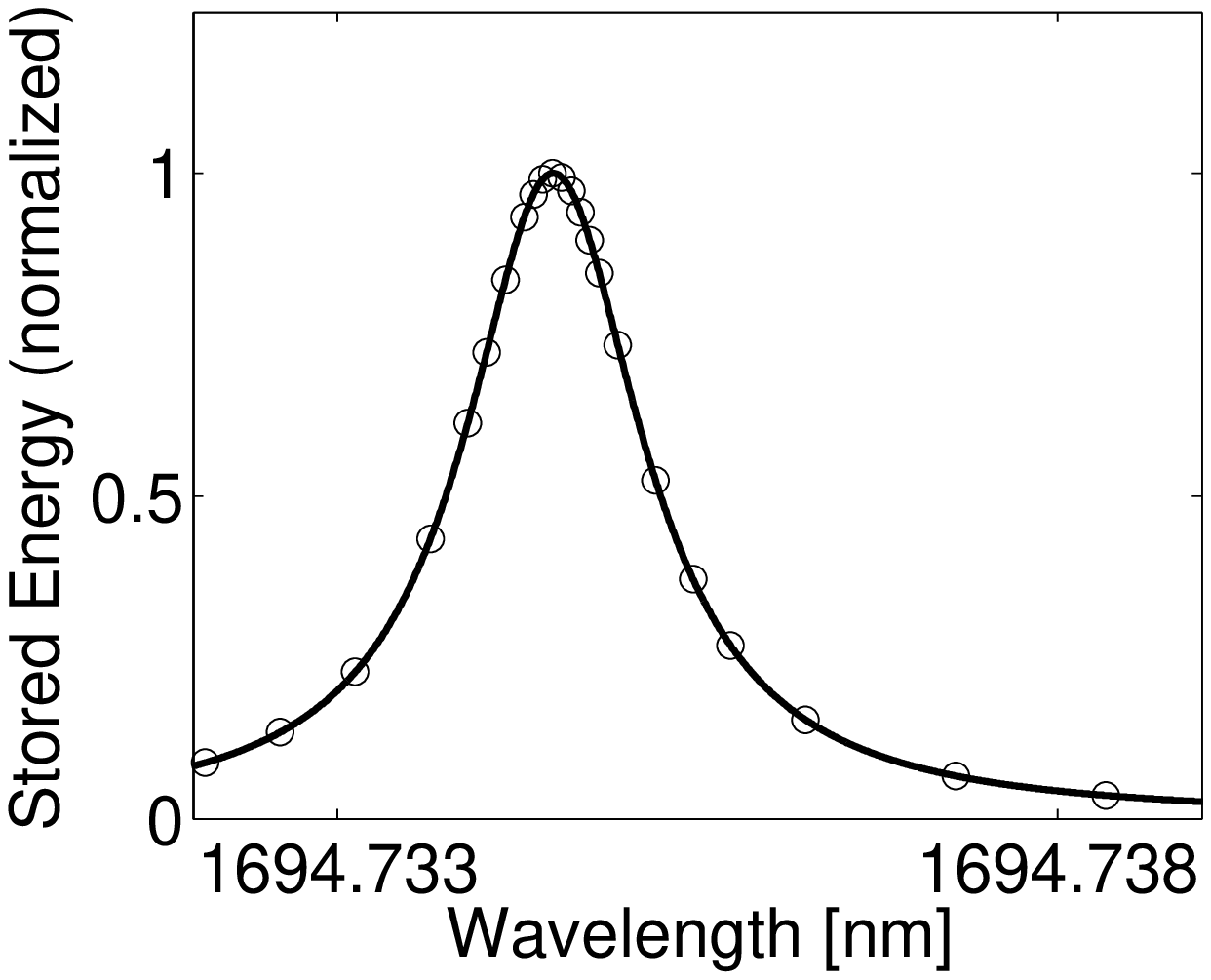}
  \includegraphics[width=.48\textwidth]{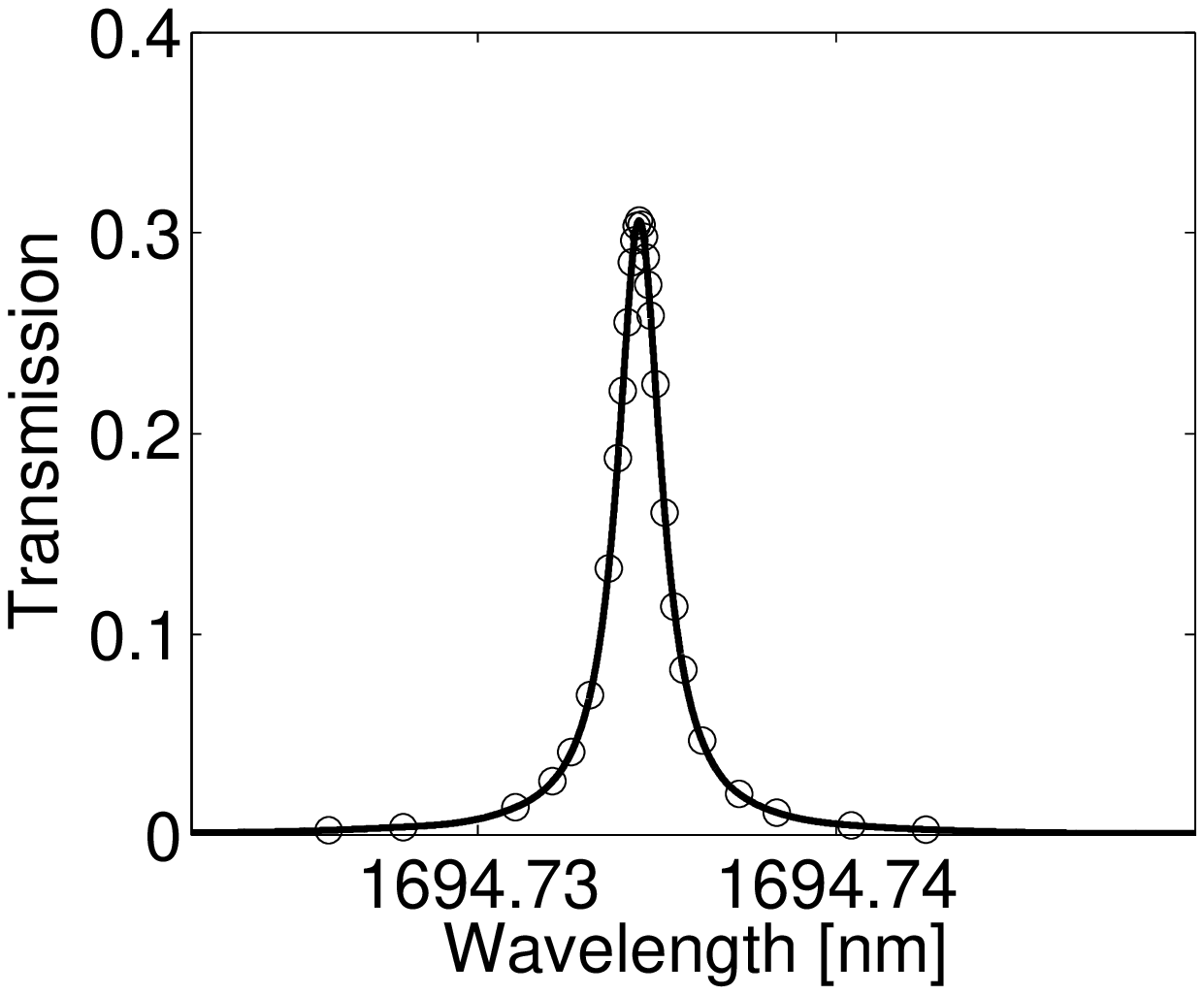}
  \caption{
Left: Energy stored in the cavity, $E$, as function of the wavelength of the incident waveguide mode, $\lambda_{in}$. 
Circles: simulation results at distinct frequencies, line: Lorentz-fit. 
Right: Transmission through the cavity, $T$, as function of $\lambda_{in}$. 
Circles: simulation results at distinct frequencies, line: spline-fit. 
}
\label{fig_resonance_scan}
\end{center}
\end{figure}

For these simulations we again investigate the modulated stack cavity with $N_i = 13$, $N_o = 5$. 
As numerical discretization parameter we choose a polynomial degree of the finite-element ansatzfunctions 
of $p=4$. 
Figure~\ref{fig_resonance_scan} (left) shows the (normalized) energy $E$ stored in the cavity as a function of the wavelength of the 
incident waveguide mode.
For automatic determination of the wavelengths of the incoming waveguide modes, $\lambda_{in}$ we use a self-adaptive 
bisection algorithm. In this example we used a search range of $\Delta \lambda\approx 3\,$nm and 14 bisections, resulting in a final 
wavelength-resolution of  $\delta \lambda\approx 10^{-4}\,$nm.
With a typical simulation time of about 10\,min for a single-wavelength simulation, this results in a total computation time 
for the frequency scan of about 8\,hrs.

We fit a Lorentzian distribution to the data points and deduce a resonance wavelength
of $\lambda_0 \approx 1694.7345\,$nm and a Q-factor of $Q\approx 1.14\times 10^6$ ($\lambda_0/FWHM$).
Both are in very good agreement with the results from the resonance mode solver 
within the ranges of numerical error (see Section~\ref{sec_resonances}), 
as expected from the convergence study.

We have also recorded the total transmission through the device by integrating the energy fluxes over the incoming and outgoing 
facets of the device. Figure~\ref{fig_resonance_scan} (right) shows the transmission peak corresponding to the resonance. 
Here the total transmission reaches a maximum of $T\approx 0.31$.

\section{Conclusion}
Size-modulated stack microcavities have been investigated numerically using 
time-harmonic finite-element solvers. 
The results have been validated by convergence analysis. 
Well converged results for cavities with Q-factors larger than $10^6$ have been obtained. 
Dependencies of resonance wavelength and quality factor on geometrical parameters have been investigated. 

\section*{Acknowledgments}
The authors would like to acknowledge the support of
European Regional Development Fund (EFRE) / Investitionsbank Berlin (IBB) through contracts 
ProFIT 10144554 and 10144555 and the support of the German Research Foundation (DFG) through 
the research center {\sc Matheon}.

{\small
\bibliography{/home/numerik/bzfburge/texte/biblios/phcbibli,/home/numerik/bzfburge/texte/biblios/group,/home/numerik/bzfburge/texte/biblios/lithography}

\begin{thebibliography}{1}

\bibitem{Vahala2003a}
Vahala, K.~J., ``Optical microcavities,'' {\em Nature}~{\bf 424},  839 -- 846
  (2003).

\bibitem{Notomi2010rep}
Notomi, M., ``Manipulating light with strongly modulated photonic crystals,''
  {\em Reports on Progress in Physics}~{\bf 73},  096501 (2010).

\bibitem{Notomi2008oe}
Notomi, M., Kuramochi, E., and Taniyama, H., ``Ultrahigh-{Q} nanocavity with 1d
  photonic gap,'' {\em Opt. Express}~{\bf 16},  11095 (2008).

\bibitem{Kuramochi2010oe}
Kuramochi, E., Taniyama, H., Tanabe, T., Kawasaki, K., Roh, Y.-G., and Notomi,
  M., ``Ultrahigh-{Q} one-dimensional photonic crystal nanocavities with
  modulated mode-gap barriers on {SiO2} claddings and on air claddings,'' {\em
  Opt. Express}~{\bf 18},  15859 (2010).

\bibitem{Joannopoulos1995a}
Joannopoulos, J.~D., Meade, R.~D., and Winn, J.~N.,  [{\em Photonic
  Crystals}{\nolinebreak\hspace{0.1em}]}, Princeton University Press,
  Princeton, NJ (1995).

\bibitem{Pomplun2007pssb}
Pomplun, J., Burger, S., Zschiedrich, L., and Schmidt, F., ``Adaptive finite
  element method for simulation of optical nano structures,'' {\em phys. stat.
  sol. (b)}~{\bf 244},  3419 (2007).

\bibitem{Burger2008ipnra}
Burger, S., Zschiedrich, L., Pomplun, J., and Schmidt, F., ``{JCMsuite}: {A}n
  adaptive {FEM} solver for precise simulations in nano-optics,'' in [{\em
  Integrated Photonics and Nanophotonics Research and
  Applications}{\nolinebreak\hspace{0.1em}]},   ITuE4, Optical Society of
  America (2008).

\bibitem{Burger2010pw3}
Burger, S., Schmidt, F., and Zschiedrich, L., ``Numerical investigation of
  photonic crystal microcavities in silicon-on-insulator waveguides,'' in [{\em
  Photonic and Phononic Crystal Materials and Devices
  X}{\nolinebreak\hspace{0.1em}]},   {\bf 7609},  76091Q, Proc. SPIE (2010).

\bibitem{Zschiedrich2006a}
Zschiedrich, L., Burger, S., Kettner, B., and Schmidt, F., ``Advanced finite
  element method for nano-resonators,'' in [{\em Physics and Simulation of
  Optoelectronic Devices XIV}{\nolinebreak\hspace{0.1em}]},   {\bf 6115},
  611515, Proc. SPIE (2006).

\end{thebibliography}
\bibliographystyle{spiebib}  
}
\end{document}